\documentclass[aip,jcp,preprint,amsmath,amssymb]{revtex4-2}
\usepackage{graphicx}
\usepackage[suffix=]{epstopdf}
\usepackage{natmove}

\newcommand{\vv}[1]{\mathbf{#1}}

\begin{document}
\title{Approximation of forces and torques from anisotropic pairwise interactions using multivariate polynomials}

\author{Mohammadreza Fakhraei}
\affiliation{Department of Chemical Engineering, Auburn University, Auburn, AL 36849, USA}

\author{Michaela Bush}
\affiliation{Department of Chemical Engineering, Auburn University, Auburn, AL 36849, USA}

\author{Chris A. Kieslich}
\email{kieslich@gatech.edu}
\affiliation{Wallace H. Coulter Department of Biomedical Engineering, Georgia Institute of Technology, Atlanta, Georgia 30332, USA}

\author{Michael P. Howard}
\email{mphoward@auburn.edu}
\affiliation{Department of Chemical Engineering, Auburn University, Auburn, AL 36849, USA}

\begin{abstract}
The dynamics of anisotropic particles are dictated by forces and torques that can be challenging to mathematically represent in computer simulations. Several data-driven approaches have been developed to approximate these interactions, but they often rely on having large amounts of training data that may be practically difficult to generate. Here, we extend a framework we recently developed for approximating anisotropic pair potentials to the approximation of pairwise forces and torques. The framework uses multivariate polynomials and physics-motivated coordinate transformations to produce accurate approximations using limited amounts of data. We first derive expressions relating the force and torque to partial derivatives of the potential energy with respect to the transformed coordinates used to represent the particle configuration. We then explore several options for approximating the forces and torques, and we critically assess their accuracy using model two- and three-dimensional shape-anisotropic nanoparticles as test cases. We find that interpolation of the pairwise potential energy produces the best result when it is known, but force and torque matching (regression) is a viable strategy when only the force and torque is available.
\end{abstract}

\maketitle

\section{Introduction}
The self-assembly of anisotropic particles is a powerful strategy for fabricating a variety of advanced materials, including colloidal supraparticles \cite{henzie:natmat:2012, yetkin:langmuir:2024, yetkin:langmuir:2025}, photonic crystals \cite{meseguer:colsurf:2005, kim:npasia:2011}, binary metamaterials \cite{elbert:sciadv:2021}, and protein nanocages \cite{chen:advsci:2023}. Particle-based simulations can play an important role in understanding and engineering  self-assembly processes, and models for the effective pairwise interactions between anisotropic particles are often needed for this purpose. However, such interactions do not typically have simple analytical functional forms, leading to great interest in approximating them using data-driven approaches \cite{nguyen:jchemphys:2022, campos-villalobos:jchemphys:2022, wilson:jchemphys:2023, argun:jchemphys:2024, lin:jchemphys:2024, hatch:jchemphys:2024, campos-villalobos:npjcompmat:2024, vinterbladh:chemarxiv:2025, argun:arxiv:2025}. These approaches have typically required large amounts of training data sampled from higher-fidelity models to achieve good accuracy, but the amount of training data that can be collected may be limited by computational cost. Better accuracy with a fixed amount of data can be achieved by leveraging physical symmetries. For example, symmetry-aware representations can be employed \cite{campos-villalobos:jchemphys:2022, lin:jchemphys:2024, campos-villalobos:npjcompmat:2024, argun:arxiv:2025}, or symmetry can be applied to reduce the extent of the configuration space \cite{argun:jchemphys:2024, hatch:jchemphys:2024, fakhraei:jpcb:2025}. However, even with symmetry accounted for, most data-driven approaches for approximating anisotropic pair potentials have still used millions of samples for training. 

We recently developed a method based on multivariate polynomial interpolation for approximating anisotropic pairwise potential energy functions from a more limited number of sampled configurations \cite{fakhraei:jpcb:2025}. The approximate pair potential $\hat{u}$ was written as an $N$-term multivariate polynomial,
\begin{equation}
\hat u(\vv{q}) = \sum_{n=0}^{N-1} c_n \psi_n(\vv{q}),
\label{eq:surrogate}
\end{equation}
where $\vv{q}$ is the relative position and orientation coordinates describing the configuration of two particles, $\psi_n$ is the $n$-th multivariate polynomial basis function, and $c_n$ is the coefficient of $\psi_n$. We used tensor products of univariate Chebyshev polynomials, as well as a combination of univariate Chebyshev and trigonometric polynomials, to construct the basis functions; these basis functions then prescribed a good set of configurations to sample. We additionally developed physics-motivated transformations to improve the effectiveness of our sampling in configuration space. We tested our approach by approximating pair potentials for two-dimensional (rod, square, and triangle) and three-dimensional (rod, cube, tetrahedron) shape-anisotropic nanoparticles comprised of particles with scaled Lennard-Jones interactions. We obtained accurate approximations of the pairwise potential energy with fewer than 2000 samples for the two-dimensional nanoparticles, fewer than 10000 samples for the three-dimensional rod, and fewer than 50000 samples for the cube and tetrahedron---a substantial reduction in training data compared to conventional machine-learning approaches \cite{nguyen:jchemphys:2022, argun:jchemphys:2024}.

In this article, we extend our framework to the approximation of the forces and torques that result from anisotropic pairwise interactions. Expressions for computing the forces and torques from the approximated potential energy are needed to perform molecular dynamics simulations, but they are nontrivial to derive under the transformations we applied. Additionally, the ability to approximate forces and torques directly, rather than indirectly through the potential energy, is important for use cases such as coarse graining \cite{nguyen:jchemphys:2022, wilson:jchemphys:2023}. Our framework for interpolating the pairwise potential energy can be trivially applied to the individual components of the Cartesian force and torque vectors; however, we show that alternative schemes produce superior approximations. Overall, we find that multivariate polynomials also approximate pairwise forces and torques well using modest amounts of training data. 

The rest of the article is organized as follows. First, we review the key points of our approach for approximating anisotropic pairwise potential energy functions using multivariate polynomials \cite{fakhraei:jpcb:2025}. Next, we derive expressions for the forces and torques for our potential-energy approximation and discuss practical considerations of its implementation. We then critically assess our framework by approximating forces and torques for the two- and three-dimensional nanoparticles we considered previously using four candidate strategies: energy interpolation, force and torque interpolation, partial derivative interpolation, and force and torque regression. We conclude with a summary of our key results and an outlook.

\section{Framework}
\subsection{Potential energy approximation}
The theoretical basis for this article is the framework we previously developed for interpolating anisotropic pair potentials using multivariate polynomials, so we first summarize its main features; the reader is referred to ref.~\citenum{fakhraei:jpcb:2025} for a complete description and justification of the procedure. The set of multivariate basis functions $\{\psi_n\}$ in eq.~\eqref{eq:surrogate} were generated from a tensor product of sets of univariate basis functions for each coordinate. The multivariate sample points (i.e., particle configurations for which the true potential energy $u$ was calculated) were generated from a tensor product of sets of univariate points that are good choices for univariate interpolation with the selected basis functions. For example, the extrema of the Chebyshev polynomials of the first kind are good sample points for those basis functions. The set of coefficients $\{c_n\}$ were then determined by solving the linear system of equations that forced the approximate potential energy $\hat u$ to interpolate $u$ at the sample points.

Anisotropic pair potentials are functions of up to six coordinates $\vv{q}$, i.e., the relative position and orientation. We represented the relative position using a scaled spherical coordinate system $\boldsymbol{\rho} = (\rho, \theta, \phi)$, where $0 \leq \rho \leq 1$ is a scaled center-to-center distance, $0 \le \theta < 2\pi$ is the azimuthal angle, and $0 \le \phi \le \pi$ is the polar angle. We defined $\rho$ as
\begin{equation}
\rho = \frac{1/r - 1/r_0}{1/(r_0+r_{\rm c}) - 1/r_0},
\label{eq:rho}
\end{equation}
where $r_0(\theta, \phi, \boldsymbol{\Omega})$ is a lower bound on the center-to-center distance $r$ that depends on both the angular position coordinates and the orientation coordinates $\boldsymbol{\Omega}$ to account for anisotropy, and $r_{\rm c}$ is a cutoff distance relative to $r_0$. We represented the relative orientation using intrinsic Euler angles $\boldsymbol{\Omega} = (\alpha,\beta,\gamma)$ defined as rotations first by $0 \le \alpha < 2\pi$ about the body-fixed $z$-axis, then by $0 \le \beta \le \pi$ about the body-fixed $x$-axis, and last by $0 \le \gamma < 2\pi$ about the body-fixed $z$-axis.

We used symmetry to reduce the approximation domain for the angular coordinates where possible. We took one particle in the pair as the origin and reduced $\theta$ and $\phi$ to the smallest values required to uniquely represent the relative position of the other particle using proper rotations. We then reduced the Euler angles to the smallest values required to uniquely represent the relative orientation of the other particle using a procedure previously developed for crystallography \cite{nolze:cryst:2015}. The domain of $\beta$ was reduced to $[0,\pi/2]$ if there was two-fold symmetry around that particle's body-fixed $x$-axis by remapping $\alpha \to \pi + \alpha$, $\beta \to \pi - \beta$, and $\gamma \to 2\pi-\gamma$ when $\beta \ge \pi/2$, and the domain of $\gamma$ was reduced by the degree of rotational symmetry about that particle's body-fixed $z$-axis using periodicity. This procedure is described here for three-dimensional particles, but it can be applied to two-dimensional particles with appropriate adaptation.

\subsection{Force and torque approximation}
\label{sec:framework:ftcalc}
Many simulation techniques, such as classical molecular dynamics \cite{allen:oxford:2017}, require the pairwise force $\vv{F} = (F_x, F_y,F_z)$ and torque $\boldsymbol{\tau} = (\tau_x, \tau_y, \tau_z)$ rather than the pairwise potential energy $u$. The force and torque are both vectors, and we will represent them in Cartesian coordinates that are convenient for simulation. Taking one particle in the pair as the origin, the force acting on the other particle is related to the derivatives of the energy with respect to its relative position $\vv{x}$ in Cartesian coordinates as $\vv{F} = -\partial_\vv{x} u$. (Throughout this discussion, $\partial_x$ denotes the partial derivative with respect to variable $x$, and $\partial_\vv{x}$ is the vector of partial derivatives with respect to the variables $\vv{x}$.) The torque on the other particle is the derivative of the potential energy with respect to the extrinsic Euler angles $\boldsymbol{\psi} = (\psi_x,\psi_y,\psi_z)$ representing an additional rotation around each of the space-fixed Cartesian axes originating at the particle's center of mass as $\boldsymbol{\tau} = -\partial_{\boldsymbol{\psi}} u$. The force and torque on the particle at the origin are then determined by Newton's third law of motion.

We did not previously calculate the forces and torques associated with our potential-energy approximation because the scaled spherical coordinates and the dependence of $\boldsymbol{\Omega}$ on $\boldsymbol{\psi}$ require nontrivial application of the chain rule to evaluate the partial derivatives with respect to $\vv{x}$ and $\boldsymbol{\psi}$. For a pairwise potential energy function $u(\vv{q})$ expressed using the coordinates $\vv{q} = (\boldsymbol{\rho}, \boldsymbol{\Omega})$, we have now derived that
\begin{equation}
\begin{bmatrix}\vv{F} \\ \boldsymbol{\tau}\end{bmatrix} = -
\vv{J}^{\rm T}\partial_{\vv{q}} u
\label{eq:force_torque}
\end{equation}
where $\vv{J}$ is a Jacobian matrix,
\begin{equation}
\vv{J} =
\begin{bmatrix}
\partial_{r}\rho & \partial_{r_0}\rho \partial_{(\theta,\phi,\boldsymbol{\Omega})}^{\rm T} r_0\\
\vv{0} & \vv{I} 
\end{bmatrix}
\begin{bmatrix}
\vv{J}_{\vv{r},\vv{x}} & \vv{0} \\
\vv{0} & \vv{J}_{\boldsymbol{\Omega},\boldsymbol{\psi}}
\end{bmatrix}
.
\end{equation}
The first 6-by-6 matrix accounts for the transformation from the conventional spherical coordinates $\vv{r} = (r,\theta,\phi)$ to our scaled spherical coordinates $\boldsymbol{\rho}$, where $\vv{I}$ is the identity matrix and $\vv{0}$ denotes zero padding. The second 6-by-6 matrix accounts for the transformation from Cartesian coordinates to spherical coordinates through the Jacobian matrix for $\vv{r}$ with respect to $\vv{x}$,
\begin{equation}
\vv{J}_{\vv{r},\vv{x}} =
\begin{bmatrix}
\cos\theta \sin\phi & \sin\theta \sin\phi & \cos\phi\\
-\sin\theta \csc\phi/r & \cos\theta\csc\phi/r & 0\\
 \cos\theta \cos\phi/r & \sin\theta\cos\phi/r & -\sin\phi/r
\end{bmatrix},
\end{equation}
as well as the dependence of the orientation coordinates on the additional rotation through the Jacobian matrix $\vv{J}_{\boldsymbol{\Omega},\boldsymbol{\psi}}$ for $\boldsymbol{\Omega}$ with respect to $\boldsymbol{\psi}$. This matrix is more involved to calculate than $\vv{J}_{\vv{r},\vv{x}}$. We did so by first adding one of the rotations $\psi_i$ in $\boldsymbol{\psi}$ to the extrinsic (space-fixed) rotation sequence corresponding to $\boldsymbol{\Omega}$. The net rotation was represented as a matrix that, except in special limiting cases (see below), can be used determine the intrinsic Euler angles for the new, rotated orientation. The chain rule was used to differentiate the new orientation angles with respect to $\psi_i$ through the relevant elements of the rotation matrix, then we evaluated the expressions when $\psi_i = 0$ to recover the derivatives in the initial orientation. Repeating this procedure for all components of $\boldsymbol{\psi}$ gave
\begin{equation}
\vv{J}_{\boldsymbol{\Omega},\boldsymbol{\psi}} =
\begin{bmatrix}
-\sin\alpha \cot\beta & \cos\alpha \cot\beta & 1\\
\cos\alpha & \sin\alpha & 0\\
\sin\alpha \csc\beta & -\cos\alpha \csc\beta & 0
\end{bmatrix}.
\label{eq:j_rot}
\end{equation}
We validated the correctness of these expressions by numerically computing $\partial_{\vv{q}} u$ for the true potential, then comparing the force and torque vectors computed using eq.~\eqref{eq:force_torque} to their directly measured values (see Sec.~\ref{sec:methods}). 

We note that $\vv{J}$ becomes undefined when $r=0$, $\phi=\{0,\pi\}$, and $\beta=\{0,\pi\}$. In practice, configurations having $r = 0$ lie outside the approximation domain because $r_0 > 0$, but configurations having these values of $\phi$ or $\beta$ readily occur. Challenges with handling these points are well-known, but we can show that the affected components of the force and torque ($F_x$, $F_y$, $\tau_x$, and $\tau_y$) in fact have well-defined limits using L'H\^{o}pital's rule. For example,
\begin{equation}
\lim_{\beta \rightarrow 0} \tau_x =  \lim_{\beta \rightarrow 0} \left[\sin\alpha( \partial_{\beta}\partial_{\alpha}u - \partial_{\beta} \partial_{\gamma} u) - \cos \alpha \ \partial_{\beta} u\right]
\label{eq:tx_singularity}
\end{equation}
is a defined limit of $\tau_x$. However, working with the second partial derivatives of the potential is cumbersome, so we propose to restrict the approximation domain for both angles to $[\delta,\pi-\delta]$, where $\delta$ is a small threshold, and use the values of the forces and torques at these bounds for the small number of configurations outside the domain. We will assess the validity of this approximation later.

In addition to calculating the force $\vv{F}$ and torque $\boldsymbol{\tau}$ from the derivatives of the potential energy $\partial_\vv{q} u$, there are also cases where it is desirable to perform the inverse operation. For example, an approximation of the potential energy can be made to best match measured forces and torques \cite{nguyen:jchemphys:2022}. This procedure can be done efficiently for an approximation having the form of eq.~\eqref{eq:surrogate} if $\vv{F}$ and $\boldsymbol{\tau}$ are first mapped to $\partial_\vv{q} u$ because determining the coefficients in eq.~\eqref{eq:surrogate} can then be formulated as a linear least squares problem \cite{lindsey:jchemtheorycomp:2017}. The required mapping can be performed by inverting eq.~\eqref{eq:force_torque} if $\vv{J}^{-1}$ exists. The determinant of $\vv{J}$ is
\begin{equation}
|\vv{J}| = \frac{r_0 \ (r_0+r_c)}{r^4 r_{\rm c} }\csc\phi \csc\beta .
\end{equation}
Hence, $\mathbf{J}$ is invertible except at the same points where it becomes undefined. We have excluded these points by the procedure described in Sec.~\ref{sec:framework:ftcalc}, so inversion of eq.~\eqref{eq:force_torque} can be performed for all points in the approximation domain. The form of the inverse matrix is complex, so we compute it numerically in practice.

\section{Methods}
\label{sec:methods}
\subsection{Anisotropic nanoparticles}
We tested our framework for approximating forces and torques on the same shapes of nanoparticles used in ref.~\citenum{fakhraei:jpcb:2025}: rod, square, and triangle in two dimensions and rod, cube, and tetrahedron in three dimensions. As in ref.~\citenum{fakhraei:jpcb:2025}, we considered only interactions between nanoparticles with the same shape for simplicity. Each nanoparticle was constructed from spherical beads of mass $m$ and diameter $\sigma$ arranged with a uniform nearest-neighbor spacing of $2\sigma/3$ and a discretization of 6 beads per axis/edge. The beads interacted with each other pairwise through a perturbed Lennard–Jones potential,\cite{weeks:jchemphys:1971}
\begin{equation} 
u_0(r) = \begin{cases} u_{\rm LJ}(r) + (1-\lambda)\varepsilon, & r \leq 2
^{1/6}\sigma \\ \lambda u_{\rm LJ}(r), & r > 2^{1/6}\sigma \end{cases},
\label{eq:true_potential} 
\end{equation} 
where $u_{\rm LJ}$ is the Lennard-Jones potential truncated and shifted to zero at 3$\sigma$, and $\varepsilon$ is the unit of energy. The total potential energy $u$ between nanoparticles is
\begin{equation}
u = \sum_{i=1}^{N_{\rm p}} \sum_{j=1}^{N_{\rm p}} u_0(|\vv{x}_{ij}|),
\end{equation}
where the sums run over the $N_{\rm p}$ particles in each nanoparticle and $\vv{x}_{ij} = \vv{x}_j - \vv{x}_i$ with $\vv{x}_i$ being the position of particle $i$. We determined $\lambda$ for each nanoparticle such that the energy for the most attractive configuration was approximately $-5\varepsilon$ by first minimizing $u$ with respect to $\vv{x}$ and $\boldsymbol{\Omega}$ when $\lambda = 1$ to identify this configuration, then using bisection search to solve for $\lambda$ that gave the desired energy. The values of $\lambda$ used are given in Table S1.

We also require the forces and torques on each nanoparticle. In the following, we label the nanoparticle associated with index $i$ as 1 and the nanoparticle associated with index $j$ as 2. The force $\vv{F}_1$ on nanoparticle 1 is
\begin{equation} 
\vv{F}_1 = \sum_{i=1}^{N_{\rm p}}\sum_{j=1}^{N_{\rm p}} \partial_{\vv{x}_i} u_0(|\vv{x}_{ij}|),
\end{equation}
and the force on nanoparticle 2 is $\vv{F}_2 = -\vv{F}_1$. The torque on nanoparticle 1 is
\begin{equation} 
\boldsymbol{\tau}_1 = \sum_{i=1}^{N_{\rm p}}\sum_{j=1}^{N_{\rm p}} (\vv{x}_i - \vv{x}_{{\rm c},1}) \times \partial_{\vv{x}_i} u_0(|\vv{x}_{ij}|)
\end{equation}
where $\vv{x}_{{\rm c},1}$ is the center of mass of nanoparticle 1, and the torque on nanoparticle 2 is $\boldsymbol{\tau}_2 = -\boldsymbol{\tau}_1 + (\vv{x}_{{\rm c},2} - \vv{x}_{{\rm c},1}) \times \vv{F}_1$.

\subsection{Multivariate polynomials}
\label{sec:methods:polynomial}
Chebyshev polynomials of the first kind and their extrema were used to construct the multivariate basis functions and sample points used for approximation. The highest-degree basis function (number of sample points) for each coordinate can be selected independently. For all the two-dimensional nanoparticles as well as the three-dimensional rod and the cube, we used the combination of number of sample points for each coordinate that gave the smallest root mean squared error in the approximation of the pairwise potential energy in our previous study \cite{fakhraei:jpcb:2025}. However, in that study, we also noted difficulty approximating the pairwise potential energy for the tetrahedron, which we speculated was due to insufficient sampling of attractive configurations with the sample points we had considered. We hence decreased the number of sample points for $\rho$ for the tetrahedron compared to ref.~\citenum{fakhraei:jpcb:2025} in order to increase the number of sample points in the other coordinates.

All sampling was performed in the reduced domain for each nanoparticle that accounts for symmetry. Using this procedure, the two-dimensional nanoparticles have three independent coordinates $(\rho, \theta, \alpha)$, the three-dimensional rod has four independent coordinates $(\rho, \phi, \alpha, \beta)$, and the cube and tetrahedron have six independent coordinates $(\rho,\theta,\phi,\alpha,\beta,\gamma)$. The reference orientation for each nanoparticle is as in ref.~\citenum{fakhraei:jpcb:2025}. Table S2 lists the number of sample points for each nanoparticle, and Table S3 lists the upper bounds of the domain for the angular coordinates. It is important to note that for the case of the three-dimensional rod, $\theta$ and $\gamma$ have been eliminated as variables, but only $\partial_\gamma u$ is zero. The rotation procedure used to eliminate $\theta$ requires $\partial_\theta u = -\partial_\alpha u$, which is in general not zero. We use this relationship when applying eq.~\eqref{eq:force_torque} for the three-dimensional rod.

We defined $r_0$ as the largest separation distance at which $u = 5\,\varepsilon$, which we evaluated using a combination of linear and bisection search. We found in ref.~\citenum{fakhraei:jpcb:2025} that it is important to have a reliable approximation of $r_0$, so we used multivariate piecewise linear interpolants on uniform grids with more sample points than we used for the interactions (Table S4). We used this approach because we found that Chebyshev polynomials did not reach the required accuracy with practical numbers of sample points, and the piecewise linear interpolants were also faster to evaluate. However, piecewise linear interpolants are not differentiable, so we estimated the required derivatives of $r_0$ using a central finite-difference scheme with step size $10^{-6}$, except at the boundaries were forward or backward differences were used instead.

We excluded the undefined points $\phi=\{0,\pi\}$ and $\beta=\{0,\pi\}$ from all our approximations by shifting the endpoints of the domain for these coordinates inward by $\delta$. To find a good value of $\delta$, we computed the $y$-components of the force and torque, $F_y$ and $\tau_y$, for $1000$ random pairs of cubes using uniform sampling of configurations with $r = r_0$, which typically had large forces and torques due to the short distance between particles, and $\phi = 0$ or $\beta = 0$ for $F_y$ or $\tau_y$, respectively, so that the relevant matrix transformations were undefined. We then computed the root mean squared error (RMSE) between the true values of $F_y$ and $\tau_y$ and those computed using eq.~\eqref{eq:force_torque} with numerically estimated partial derivatives for the same configuration but with $\phi=\delta$ for $F_y$ or $\beta=\delta$ for $\tau_y$. A too-large value of $\delta$ will give a poor approximation of the limiting value at the endpoint, but a too-small value of $\delta$ may cause numerical issues. We confirmed this behavior in the RMSE for both $F_y$ and $\tau_y$ as a function of $\delta$ (Fig.~S1), and we chose to use $\delta=10^{-5}$ because it gave the smallest RMSE for the values considered.

\section{Results and Discussion}
\begin{figure*}
    \includegraphics{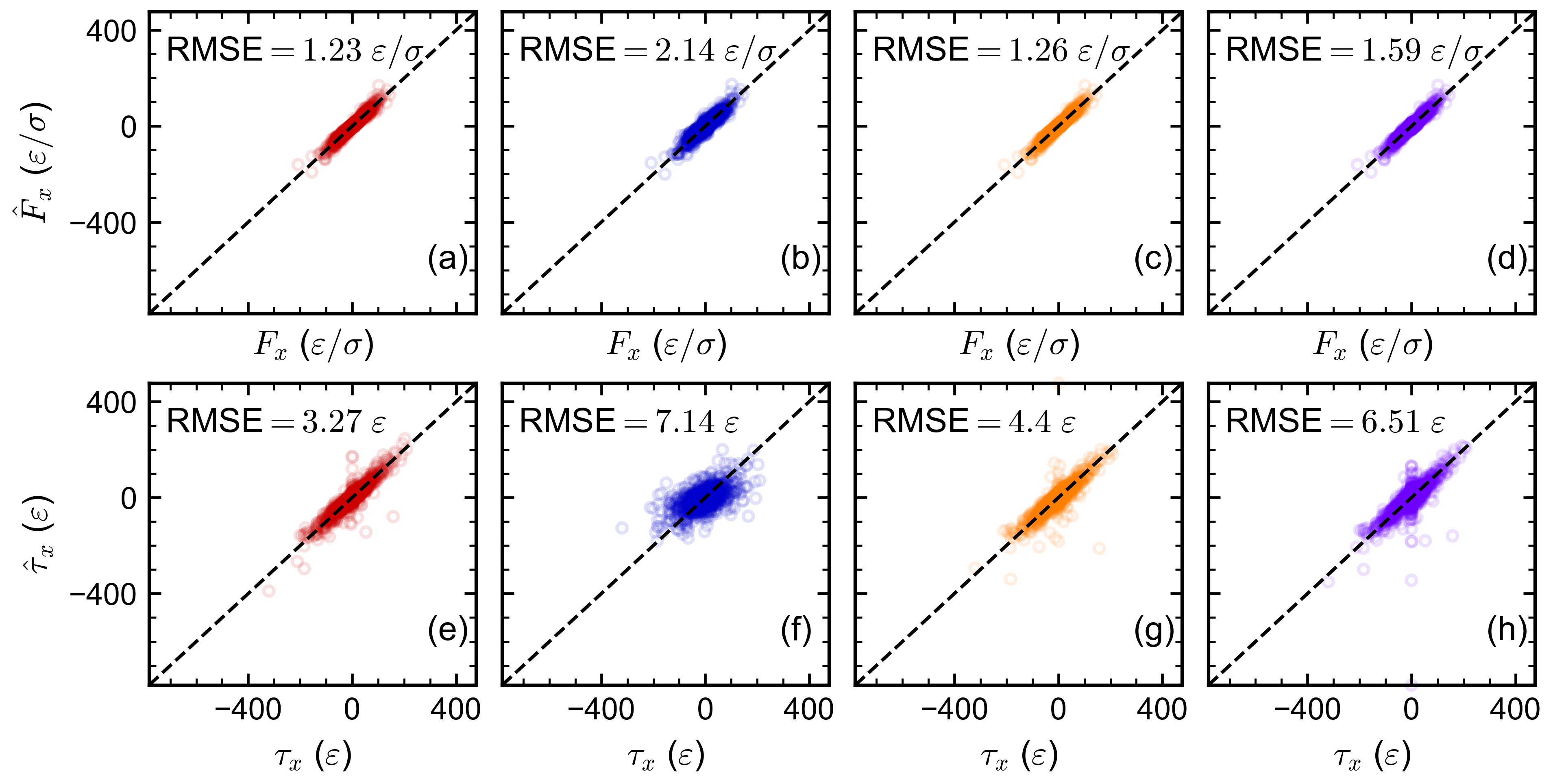} 
    \caption{Approximated $x$-component of the (a--d) force $\hat{F}_x$ and (e--h) torque $\hat{\tau}_x$ vs.~true values $F_x$ and $\tau_x$ for the cube. The approximations were constructed by (a,e) interpolation of energy (E), (b,f) interpolation of forces and torques (FT), (c,g) interpolation of partial derivatives (D), and (d,h) regression of forces and torques (FT-R). To facilitate comparison, all panels share the same axis limits, which are set by the minimum and maximum of all true and approximated values. The root mean squared error (RMSE) for each approximation is also stated.}
    \label{fig:cube_parity_main}
\end{figure*}

\begin{figure*}
    \includegraphics{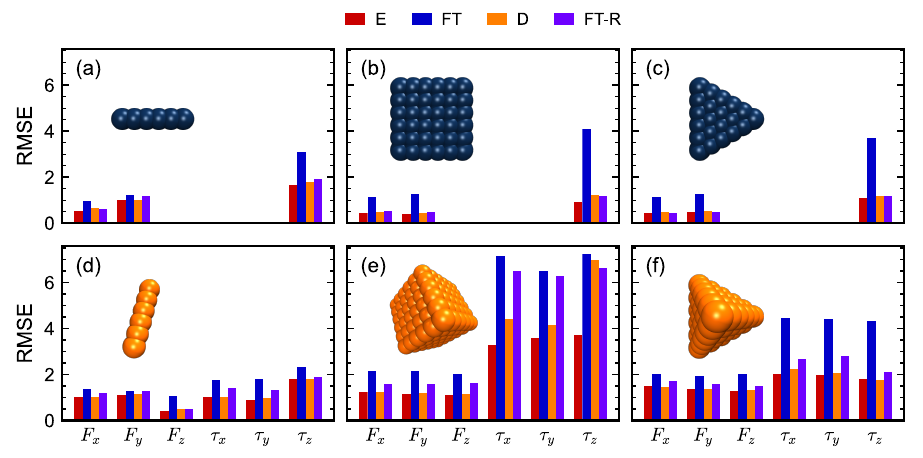} 
    \caption{RMSE for all force and torque components for (a) two-dimensional rod, (b) square, (c) triangle, (d) three-dimensional rod, (e) cube, and (f) tetrahedron approximated using the different strategies (see Fig.~\ref{fig:cube_parity_main}). The unit of RMSE is $\varepsilon/\sigma$ for the force and $\varepsilon$ for the torque. Corresponding parity plots for all components of the force and torque for all nanoparticles are shown in Figs.~S2--S7. Nanoparticle images were visualized using VMD 1.9.4 \cite{humphrey:jmolgrp:1996}.}
    \label{fig:error}
\end{figure*}

We first tested the ability of approximations of the pairwise potential energy $\hat{u}$ like those we developed previously\cite{fakhraei:jpcb:2025} to predict forces and torques. Specifically, $\hat{u}$ was constructed to interpolate the energy measured at the sample points. We then generated a set of uniformly random configurations within the approximation domain for each nanoparticle ($10^4$ for the two-dimensional nanoparticles, $5\times10^4$ for the three-dimensional nanoparticles) to use for testing. We calculated the approximate force $\vv{\hat{F}}$ and torque $\boldsymbol{\hat{\tau}}$ from $\hat u$ for each of these configurations using eq.~\eqref{eq:force_torque}, then we determined the root mean squared error (RMSE) for each component of these vectors relative to the true force $\vv{F}$ and torque $\boldsymbol{\tau}$ across all configurations. Figure \ref{fig:cube_parity_main} shows parity plots for the $x$-components of the force and torque for the cube nanoparticle computed by different approximations. The approximated force $\hat F_x$ and torque $\hat\tau_x$ computed by interpolating the potential energy [Figs.~\ref{fig:cube_parity_main}(a) and \ref{fig:cube_parity_main}(e)] were in excellent agreement with the true values $F_x$ and $\tau_x$, having RMSEs that were only 0.35\% of the range of true values for the force and 0.62\% for the torque. Figure \ref{fig:error} summarizes the RMSEs for all approximated force and torque components across all nanoparticles; comparably small RMSE was obtained in all cases using interpolation of the energy. We do note that the RMSE for the torque was generally somewhat larger than for the force, despite both having comparable ranges and magnitudes. The larger RMSE may suggest that better approximation is needed with respect to the orientational coordinates, for example, with more sampling or different basis functions \cite{fakhraei:jpcb:2025}. 

We next turned to the challenge of approximating measurements of the force and torque rather than the energy, which arises in the context of coarse graining \cite{noid:jcp:2008, nguyen:jchemphys:2022}. One possible strategy is to approximate each component of the force and torque independently \cite{argun:jchemphys:2024}, i.e., build approximations having the same form as eq.~\eqref{eq:surrogate} but now for each component of $\vv{\hat{F}}$ and $\boldsymbol{\hat{\tau}}$. It is simple both to construct the approximations and to evaluate the forces and torques because no transformations like eq.~\eqref{eq:force_torque} are required. The independent approximations are also guaranteed to interpolate all measured forces and torques. Interestingly, though, interpolations of $\hat{F}_x$ [Fig.~\ref{fig:cube_parity_main}(b)] and $\hat{\tau}_x$ [Fig.~\ref{fig:cube_parity_main}(f)] for the cube were less accurate than interpolation of the energy, and larger RMSE was found for all nanoparticles and all vector components (Fig.~\ref{fig:error}) using this strategy.

\begin{figure}
    \includegraphics{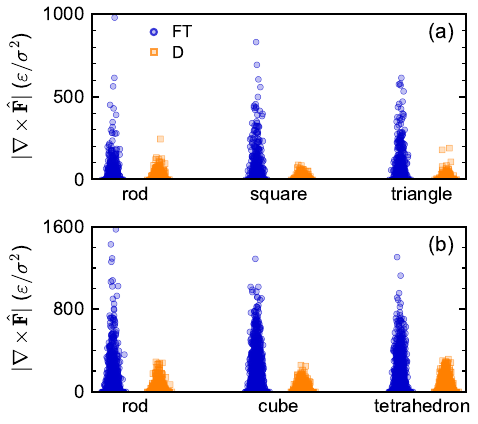} 
    \caption{The magnitude of the curl of the approximated forces $\vv{\hat{F}}$ using interpolation of forces and torques (FT) or interpolation of partial derivatives (D) for the (a) two-dimensional and (b) three-dimensional nanoparticles. The curl of the force is zero for both interpolation of energy (E) and regression of forces and torques (FT-R).}
    \label{fig:curl}
\end{figure}

This result may seem counterintuitive because the quantities of interest are more poorly approximated when they are interpolated directly rather than calculated indirectly. One possible explanation may be that the energy has features that are easier to represent using the chosen basis functions than the forces and torques that derive from it have. Another possible explanation may be that interpolating the energy produces better approximations because the components of the force and torque vectors are mathematically related to each other and not in fact independent. For example, the curl of a conservative force must be zero, but we found that it was not using force and torque interpolation (Fig.~\ref{fig:curl}). These relationships are encoded to some extent in the energy and enforced when the forces and torques are calculated from an approximation of it, but they are not when each component is interpolated independently.

We were curious whether better approximations might be produced by interpolating the partial derivatives of the energy with respect to the transformed coordinates $\vv{q}$ rather than the Cartesian force and torque vectors. The partial derivatives at the sample points were computed by inverting eq.~\eqref{eq:force_torque} for the measured forces and torques, then we interpolated each partial derivative. We evaluated the approximated partial derivatives at the test points and used eq.~\eqref{eq:force_torque} to recover the corresponding force and torque. This strategy produced a superior approximation of $\hat{F}_x$ [Fig.~\ref{fig:cube_parity_main}(c)] and $\hat{\tau}_x$ [Fig.~\ref{fig:cube_parity_main}(g)] for the cube than interpolation of the forces and torques; however, the interpolation of energy was still better. This trend was consistent across the other components of the force and torque and the other nanoparticles (Fig.~\ref{fig:error}). We note that the interpolation of partial derivatives produced a few outlier points for which $\hat{\tau}_x$ differed substantially from its true value [Fig.~\ref{fig:cube_parity_main}(g)], and these points contributed significantly to the RMSE. The curl of the approximated force was also closer to zero using the interpolated partial derivatives (Fig.~\ref{fig:curl} and Table S5). Hence, we considered interpolation of the partial derivatives to perform better than interpolation of forces and torques. Possible reasons for this improvement could be that the partial derivatives with respect to $\vv{q}$ may be easier to approximate due to the coordinate transformation, that transformation between partial derivatives and forces and torques helps encode some physics that no longer needs to be approximated, and that the transformation process effectively mixes information between vector components so each is no longer treated independently.

Finally, we considered a strategy that fits a single approximation of the energy $\hat{u}$ to its partial derivatives at the sample points, which are obtained from the measured forces and torques, rather than independently interpolating each. This approach guarantees that the approximated forces and torques are consistent with each other (e.g., the curl of the force is zero). The coefficients of $\hat{u}$ were requested to satisfy an overdetermined system of linear equations to match each partial derivative at each sample point, which we solved in a least-squares sense. The approximated forces and torques need not match all measured values because they are being regressed rather than interpolated, but they must be consistent because they derive from the same energy function. This strategy is equivalent to force and torque matching in the context of coarse graining \cite{nguyen:jchemphys:2022}. We approximated $\hat{u}$ using this strategy and evaluated the forces and torques from it; they had an accuracy intermediate between the interpolation of forces and torques and interpolation of partial derivatives [Figs.~\ref{fig:cube_parity_main}(d), \ref{fig:cube_parity_main}(h), and \ref{fig:error}]. The RMSE from force and torque regression was typically closer to interpolation of the partial derivatives, except in the case of the cube and particularly its torque. This difference for the cube may be due to the sampling design, which used the least number of points for the angular components (Table S2).

In summary, the four strategies were ranked from greatest to least based on RMSE in the approximated forces and torques as: interpolation of energy (E), interpolation of partial derivatives (D), regression of forces and torques (FT-R), and interpolation of forces and torques (FT). However, in selecting between strategies, it may also be important to consider accuracy for approximating the energy, which ultimately dictates the equilibrium thermodynamics. This consideration is especially important for the strategies that did not use the energy for fitting (D, FT, and FT-R). Unfortunately, it is not possible to directly compare the energy across all strategies because FT and D are nonconservative (i.e., the forces and torques do not derive from a single energy function) and it is not obvious what level of error is acceptable even for the two cases where the energy is known (E and FT-R).

Instead, we performed constant-temperature molecular dynamics simulations of two three-dimensional rods using our different strategies, and we measured the distribution of pairwise distances that resulted. One rod had its center of mass fixed at the origin in its reference orientation, while the other moved around it. The moving rod was confined by a spherical barrier at radial distance $8\,\sigma$ from the origin, which we represented using the repulsive half of a harmonic potential with spring constant $100\,\varepsilon/\sigma^2$, so that it remained near the fixed rod. The simulations were performed with HOOMD-blue \cite{anderson:compmatsci:2020} (version 5.4.0) and azplugins \cite{azplugins} (version 1.1.0) using a weak Langevin thermostat with translational friction coefficient $0.1\,m/\tau$, isotropic rotational friction coefficient $0.1\,\tau^{-1}$, and temperature $1\,\varepsilon/k_{\rm B}$ where $\tau=\sqrt{m\sigma^2/\varepsilon}$ is the unit of time and $k_{\rm B}$ is the Boltzmann constant. The simulation timestep was $0.005\,\tau$.

\begin{figure}
    \includegraphics{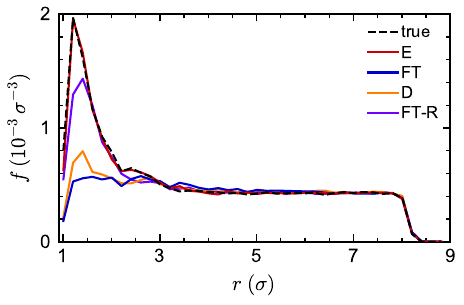} 
    \caption{Probability density function $f$ to find one three-dimensional rod a distance $r$ from the center of mass of a fixed rod. The dashed black line is the equilibrium distribution for the true interactions. The other lines are the distribution after simulating an additional $500\,\tau$ using the four different approximation strategies (see Fig.~\ref{fig:cube_parity_main}). The normalization is such that $\int {\rm d}{r}\,4\pi r^2 f(r) = 1$.}
    \label{fig:probability}
\end{figure}

We first simulated rods that interacted through the true pair potential using an explicit representation of the rod's constitutent particles and eq.~\eqref{eq:true_potential}. We performed an ensemble of $10^5$ independent simulations. The initial position and orientation of the moving rod were drawn using uniformly random sampling with $r_0 \le r \le 8\,\sigma$, while its initial linear and angular momentum were drawn from the relevant Boltzmann distribution. We simulated for $5000\,\tau$ and recorded the final configuration of the moving rod in each simulation. We then computed the probability density function $f(r)$ to observe the free rod at a radial distance $r$ from the origin (center of mass of the fixed rod) using a histogram with bin width $0.2\,\sigma$ of all configurations (dashed line in Fig.~\ref{fig:probability}). We confirmed that this simulation time was sufficiently long that the distribution did not change and so can be considered the true equilibrium distribution.

After generating an equilibrated ensemble of configurations, we then ran simulations using the four different strategies for approximating the force and torque to characterize how the distribution of pairwise distances changed (Fig.~\ref{fig:probability}). The simulation time was restricted to $500\,\tau$ due to the slow performance of our prototype implementation of the force and torque approximations; we comment on this limitation more in Sec.~\ref{sec:conclusions}. Accordingly, we started the moving rod from the final configurations obtained using the true potential so we did not need to simulate the equilibration process, and we measured departures from the true distribution during the simulation. We confirmed that these simulations were long enough to make this assessment by comparing with the distributions at shorter times, finding they changed only negligibly.

A perfect approximation of the force and torque should produce the same distribution as the simulations using the true interactions. Consistent with their RMSEs for the force and torque, E gave the best agreement with the true distribution (essentially quantitative), while FT gave the worst agreement. Interestingly, FT-R was in better agreement with the true distribution than D, despite FT-R having a larger RMSE than D (Fig.~\ref{fig:error}). Equilibrium configurational distributions should be determined by the potential energy, so this discrepancy is most likely due to the nonconservative nature of the D strategy (Fig.~\ref{fig:curl}). It also emphasizes the importance of considering not only the accuracy of an approximation for its direct outputs but also relationships with other quantities when assessing data-driven models for particle interactions.

\section{Conclusions}
\label{sec:conclusions}
In this work, we have extended our framework for approximating anisotropic pair potentials using multivariate polynomials to the approximation of pairwise forces and torques. We first derived expressions relating the forces and torques to partial derivatives of the potential energy with respect to the coordinates that we previously showed help produce good approximations of the energy using only a small number of samples \cite{fakhraei:jpcb:2025}. We then tested four strategies for approximating forces and torques using tensor products of Chebyshev polynomials of the first kind: interpolation of energy (E), interpolation of forces and torques (FT), interpolation of partial derivatives (D), and regression of forces and torques (FT-R). E requires computing the energy at the sample points but FT, D, and FT-R require only forces and torques at the sample points, which is attractive for applications such as coarse graining\cite{nguyen:jchemphys:2022}. FT-R differs conceptually from E, FT, and D because it is based on regression rather than interpolation and so need not produce an approximation that matches all sampled data.

For a set of model two- and three-dimensional nanoparticles, E performed best, FT performed worst, D performed better than FT and was closer to the accuracy of E, while FT-R typically performed somewhere between D and FT. We noted that both E and FT-R parametrize a single potential energy function, which guarantees consistency between the approximated forces and torques, whereas FT and D do not. As a practical consequence, FT and D both gave nonconservative forces, contrary to the true model, and FT-R approximated the distribution of pairwise distance for two three-dimensional rods better than D even though D gave a better approximation of the individual components of the force and torque. We hence recommend interpolation of the energy (E) when possible, and otherwise, regression of the forces and torques (FT-R). For the latter strategy, regularization (e.g., ridge\cite{hoerl:technometrics:1970} or lasso\cite{tibshirani:jroyalstat:1996} approaches) may be helpful to mitigate overfitting and produce a smoother potential-energy surface \cite{lindsey:jchemtheorycomp:2017, lindsey:jchemtheorycomp:2019, pham:jchemphys:2020, lindsey:jchemphys:2020, lindsey:jchemphys-2:2020, goldman:jchemphys:2023, lindsey:jchemphys:2023, lindsey:natmat:2025}. 

In ref.~\citenum{fakhraei:jpcb:2025}, we found there were possible benefits to using a mixture of both Chebyshev and trigonometric polynomials to approximate the potential energy and its partial derivatives, rather than only Chebyshev polynomials as used here. Specifically, we proposed to use the trigonometric polynomials for coordinates that are periodic. We speculate that these benefits may also extend to approximating forces and torques because they are related to partial derivatives of the energy. However, we chose not to use the mixed-basis polynomials in this study for computational convenience: the linear least-squares regression for FT-R was more challenging for the mixed-basis polynomials because it involved complex values and hence doubled memory requirements. This computational difficulty might be overcome using a distributed computing approach \cite{lindsey:jchemphys-2:2020} or alternative fitting procedure. 

We have currently implemented only a Python-based prototype of our multivariate-polynomial approximations. With this implementation, it was feasible for us to simulate three-dimensional rods but not the other three-dimensional nanoparticles, which were described by more coordinates and approximated using more terms, and so were more expensive to evaluate. We emphasize that we do not believe this to be an actual computational limitation of our approach. For example, multibody interactions represented using Chebyshev polynomials have previously been simulated successfully at large scale \cite{lindsey:jchemtheorycomp:2017, lindsey:jchemtheorycomp:2019,pham:jchemphys:2020, lindsey:jchemphys:2020, lindsey:jchemphys-2:2020, goldman:jchemphys:2023,lindsey:jchemphys:2023, lindsey:natmat:2025}. We intend to develop a highly performant implementation of our multivariate-polynomial approximations for performing simulations soon.

\section*{Supplementary Material}
See the supplementary material for the perturbed Lennard-Jones interaction parameter $\lambda$, approximation details, parity plots of forces and torques, and statistics related to Fig.~\ref{fig:curl} for all nanoparticles.

\section*{Conflicts of interest}
The authors have no conflicts to disclose.

\section*{Data Availability}
The data that support the findings of this study are available from the authors upon reasonable request.

\section*{Acknowledgments}
We acknowledge support from the Auburn University Research Support Program (M.F.), the National Institutes of Health under Award No.~R35GM147164 (C.A.K.), and the National Science Foundation under Award No.~2223084 (M.P.H). This work was completed with resources provided by the Auburn University Easley Cluster.

\bibliography{references}

\end{document}


\title{Supplementary material for ``Approximation of forces and torques from anisotropic pairwise interactions using multivariate polynomials''}

\author{Mohammadreza Fakhraei}
\affiliation{Department of Chemical Engineering, Auburn University, Auburn, AL 36849, USA}

\author{Michaela Bush}
\affiliation{Department of Chemical Engineering, Auburn University, Auburn, AL 36849, USA}

\author{Chris A. Kieslich}
\email{kieslich@gatech.edu}
\affiliation{Wallace H. Coulter Department of Biomedical Engineering, Georgia Institute of Technology, Atlanta, Georgia 30332, USA}

\author{Michael P. Howard}
\email{mphoward@auburn.edu}
\affiliation{Department of Chemical Engineering, Auburn University, Auburn, AL 36849, USA}

\maketitle
\begin{table}[!h]
    \caption{Perturbed Lennard-Jones potential [eq.~(9)] interaction parameter $\lambda$.}
    \begin{tabularx}{3in}{CC}
    nanoparticle & $\lambda$\\
    \hline
    rod (2D) & 0.363\\
    square & 0.279 \\
    triangle & 0.265\\
    rod (3D) & 0.363\\
    cube & 0.021\\
    tetrahedron & 0.031
    \end{tabularx}
\end{table}

\begin{table}[!h]
    \caption{Number of sample points for each coordinate for interaction approximations.}
    \begin{tabularx}{5in}{CCCCCCC}
    nanoparticle & $\rho$ & $\theta$ & $\phi$ & $\alpha$ & $\beta$ & $\gamma$\\
    \hline
    rod (2D) & 17 & 9 & - & 9 & - & - \\
    square & 17 & 9 & - & 9 & - & -\\
    triangle & 17 & 5 & - & 17 & - & - \\
    rod (3D) & 17 & - & 5 & 17 & 5 & - \\
    cube & 17 & 3 & 5 & 17 & 3 & 3 \\
    tetrahedron & 10 & 4 & 7 & 10 & 7 & 4
    \end{tabularx}
    \label{tab:points_combination}
\end{table}

\begin{table}[!h]
    \caption{Upper bounds of position angles and Euler angles.}
    \begin{tabularx}{4.5in}{CCCCCC}
    nanoparticle & $\theta$ & $\phi$ & $\alpha$ & $\beta$ & $\gamma$\\
    \hline
    rod (2D) & $\pi/2$ & - & $\pi$ & - & - \\
    square & $\pi/4$ & - & $\pi/2$ & - & - \\
    triangle & $\pi/3$ & - & $2\pi/3$ & - & - \\
    rod (3D) & - & $\pi/2$ & $2\pi$ & $\pi/2$ & - \\
    cube & $\pi/4$ & $\pi/2$ & $2\pi$ & $\cos^{-1}(1/\sqrt{3})$ & $\pi/2$\\
    tetrahedron & $2\pi/3$ & \ $\pi-10^{-5}$ & $2\pi$ & $\pi-10^{-5}$ & $2\pi/3$
    \end{tabularx}
    \label{tab:angles}
\end{table}

\begin{table}[!h]
    \caption{Number of sample points for each coordinate for $r_0$ approximations.}
    \begin{tabularx}{4.5in}{CCCCCC}
    nanoparticle & $\theta$ & $\phi$ & $\alpha$ & $\beta$ & $\gamma$\\
    \hline
    rod (2D) & 65 & - & 129 & - & - \\
    square & 65 & - & 129 & - & - \\
    triangle & 65 & - & 129 & - & - \\
    rod (3D) & - & 33 & 129 & 33 & - \\
    cube & 17 & 33 & 65 & 17 & 33\\
    tetrahedron & 17 & 33 & 65 & 33 & 17
    \end{tabularx}
\end{table}

\begin{figure}[!h]
    \includegraphics{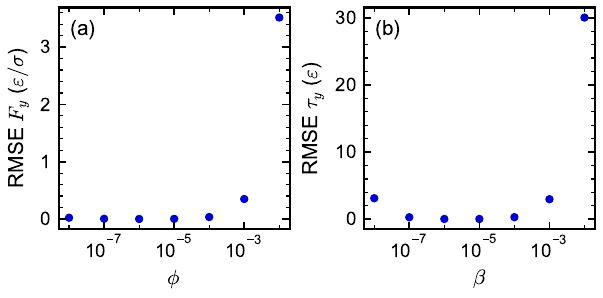} 
    \caption{RMSE for calculation of (a) $F_y$ and (b) $\tau_y$ for a cube near endpoint of $\phi$ and $\beta$. The partial derivatives of the potential energy were computed numerically at each value of $\phi$ or $\beta$, then the force and torque were calculated using eq.~(3). These values were compared to the true value of the force and torque when $\phi = 0$ or $\beta = 0$.}
\end{figure}

\begin{figure}[!h]
    \centering
    \rotatebox{90}{
      \begin{minipage}{\linewidth}
        \centering
        \includegraphics[width=\linewidth]{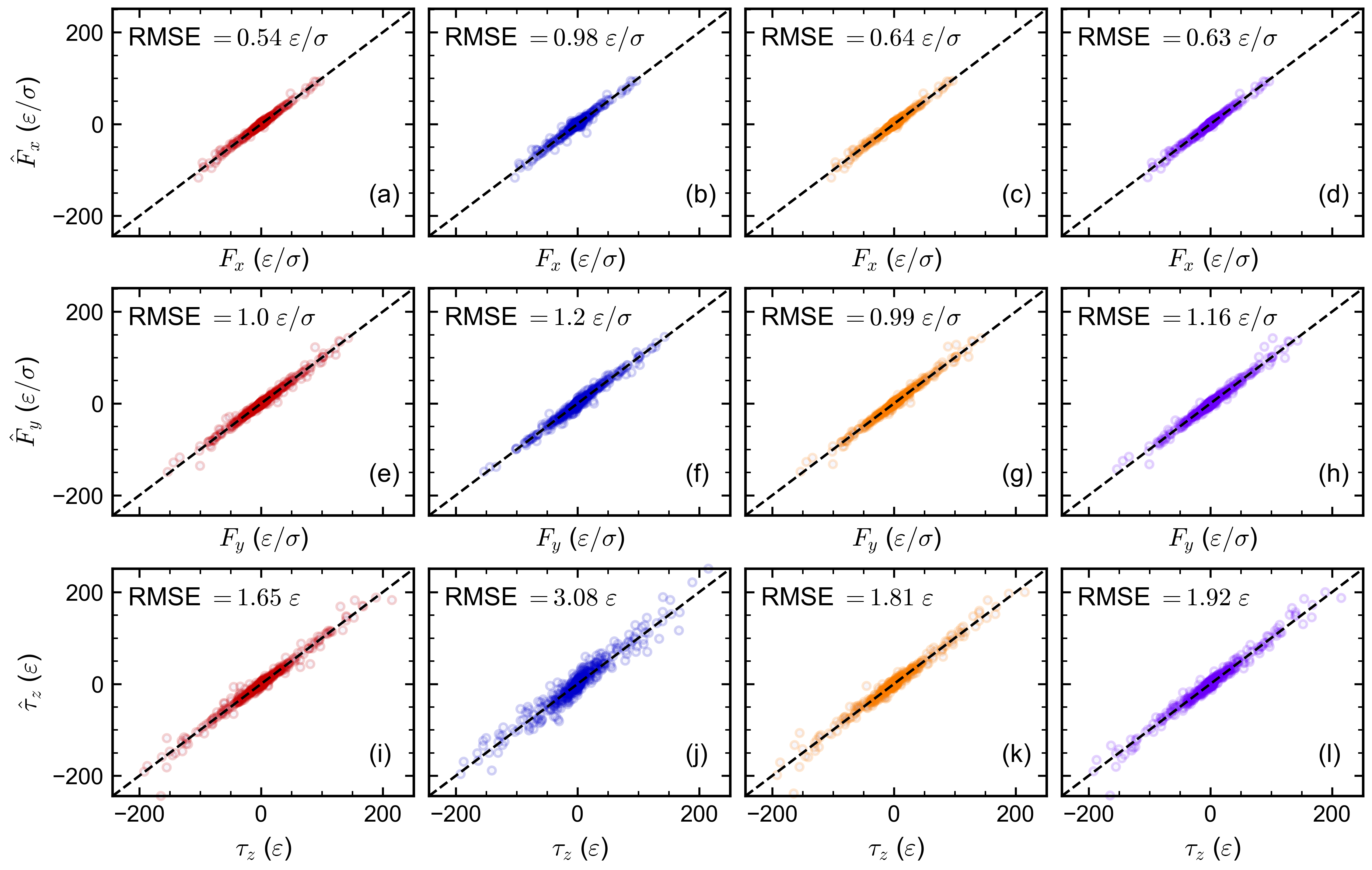}
        \caption{Parity plots of approximations of (a--d) $x$-component of force $\hat{F}_x$, (e--h) $y$-component of force $\hat{F}_y$, and $z$-component of torque $\hat{\tau}_z$ vs.~true values for the two-dimensional rod.  Each column is a different approximation strategy, as in Fig.~1 (left to right): interpolation of energy (E), interpolation of force and torque (FT), interpolation of partial derivatives (D), and regression of force and torque (FT-R). \label{fig:rod2d_parity}}
      \end{minipage}
    }
\end{figure}

\begin{figure}[!h]
    \centering
    \rotatebox{90}{
      \begin{minipage}{\linewidth}
        \centering
        \includegraphics[width=\linewidth]{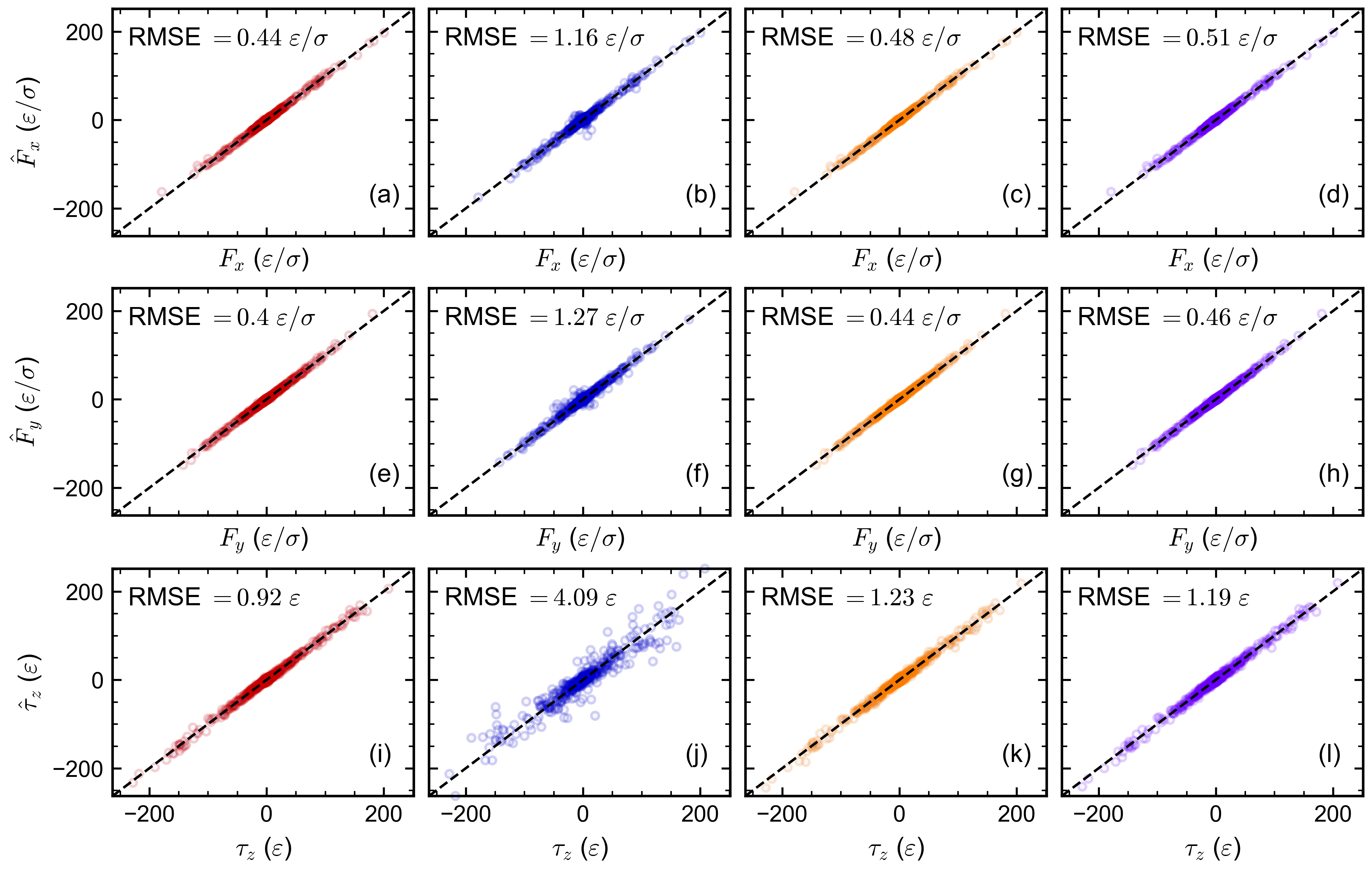}
        \caption{Same as Fig.~\ref{fig:rod2d_parity} but for the square.}
      \end{minipage}
    }
\end{figure}

\begin{figure}[!h]
    \centering
    \rotatebox{90}{
      \begin{minipage}{\linewidth}
        \centering
        \includegraphics[width=\linewidth]{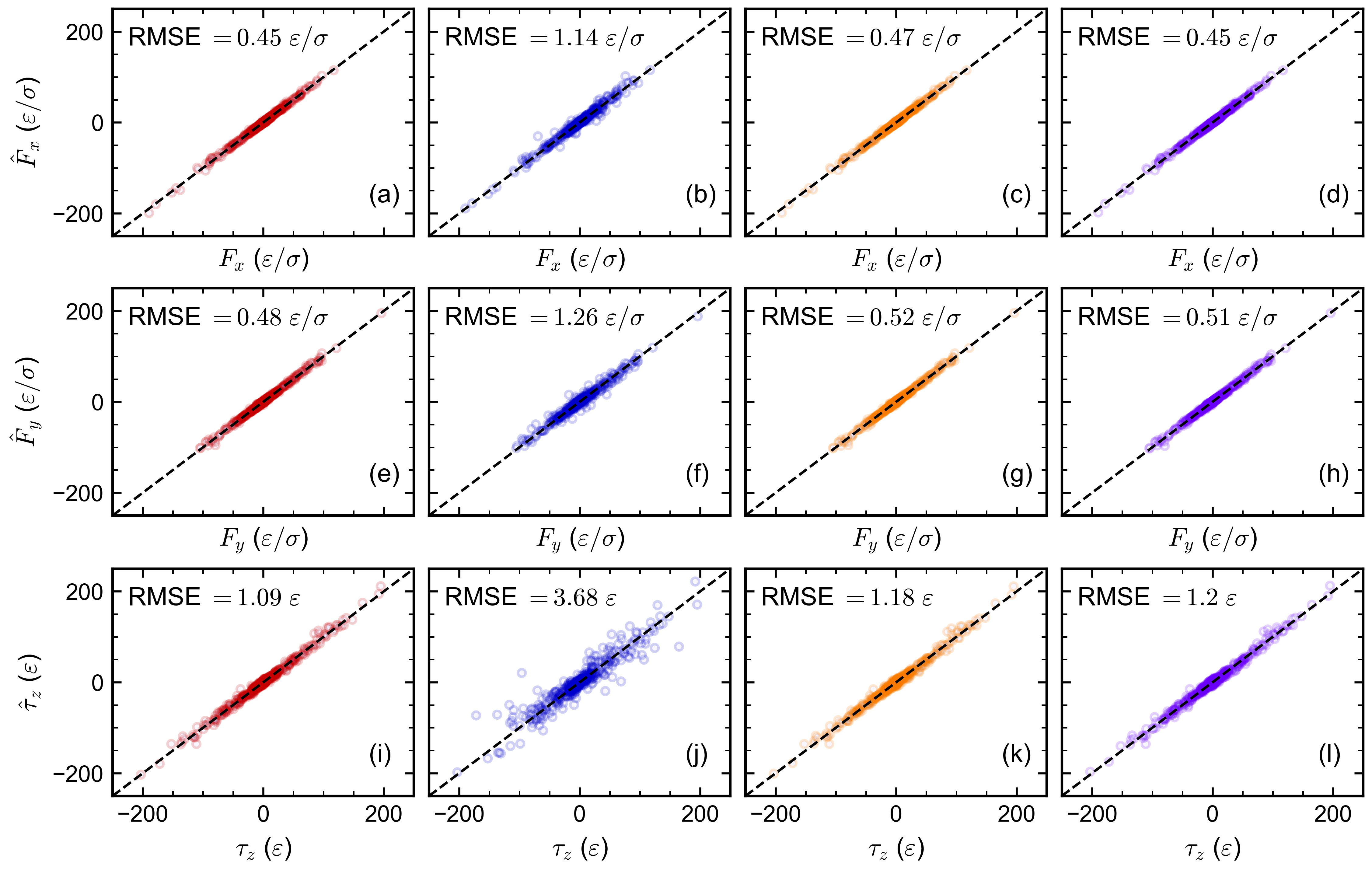}
        \caption{Same as Fig.~\ref{fig:rod2d_parity} but for the triangle.}
      \end{minipage}
    }
\end{figure}

\begin{figure}[!h]
        \includegraphics[width=\linewidth]{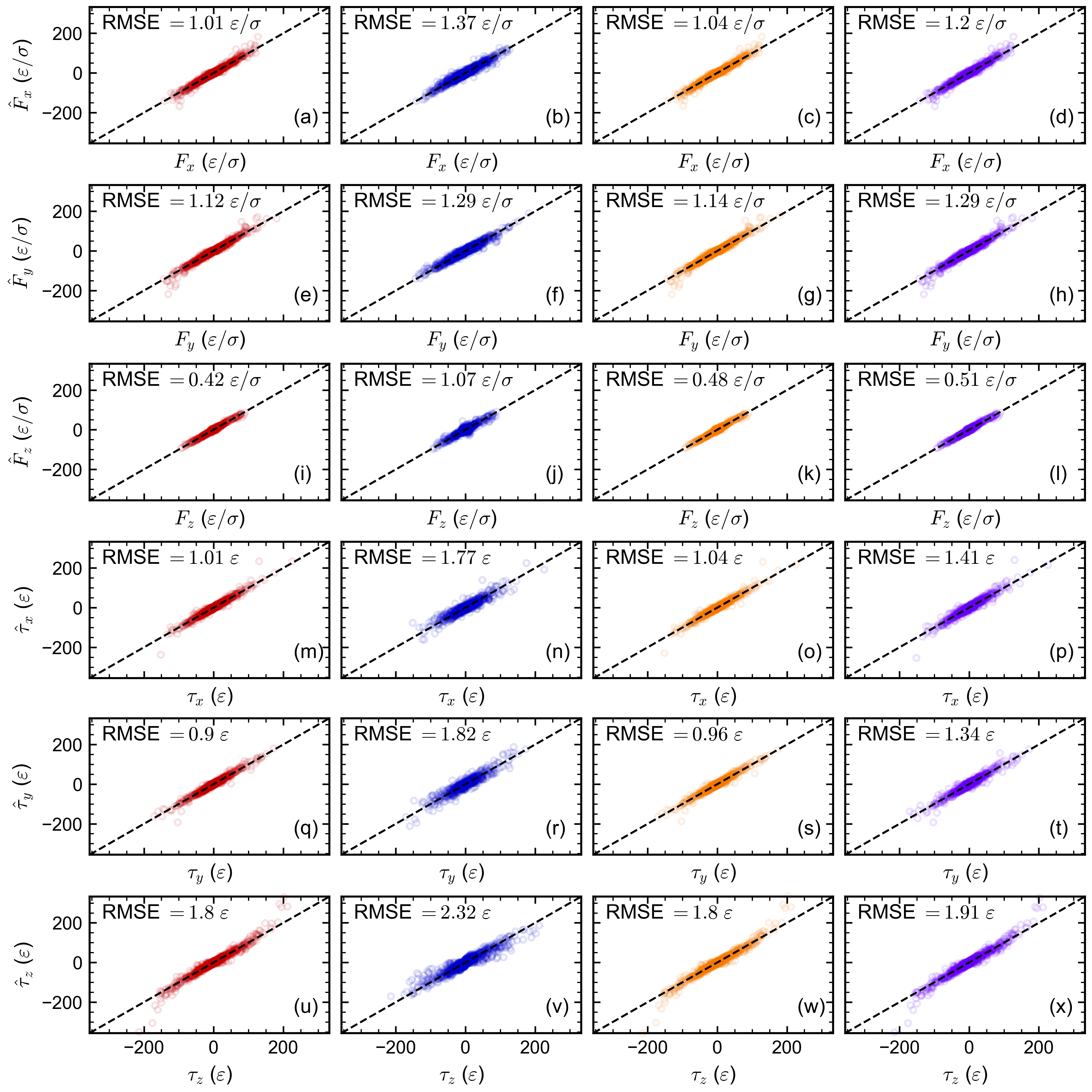}
        \caption{Parity plots of approximations of (a--d) $x$-component of force vector $\hat{F}_x$ , (e--h) $y$-component of the force vector $\hat{F}_y$ , (i--l) $z$-component of the force vector $\hat{F}_z$ , (m--p) $x$-component of force torque vector $\hat{\tau}_x$ , (q--t) $y$-component of the torque vector $\hat{\tau}_y$ , and (u--x) $z$-component of the torque vector $\hat{\tau}_z$ vs.~true values for the three-dimensional rod. Each column is a different approximation strategy as described in Fig.~\ref{fig:rod2d_parity}}
    \label{fig:rod3d_parity}
\end{figure}

\begin{figure}[!h]
        \includegraphics[width=\linewidth]{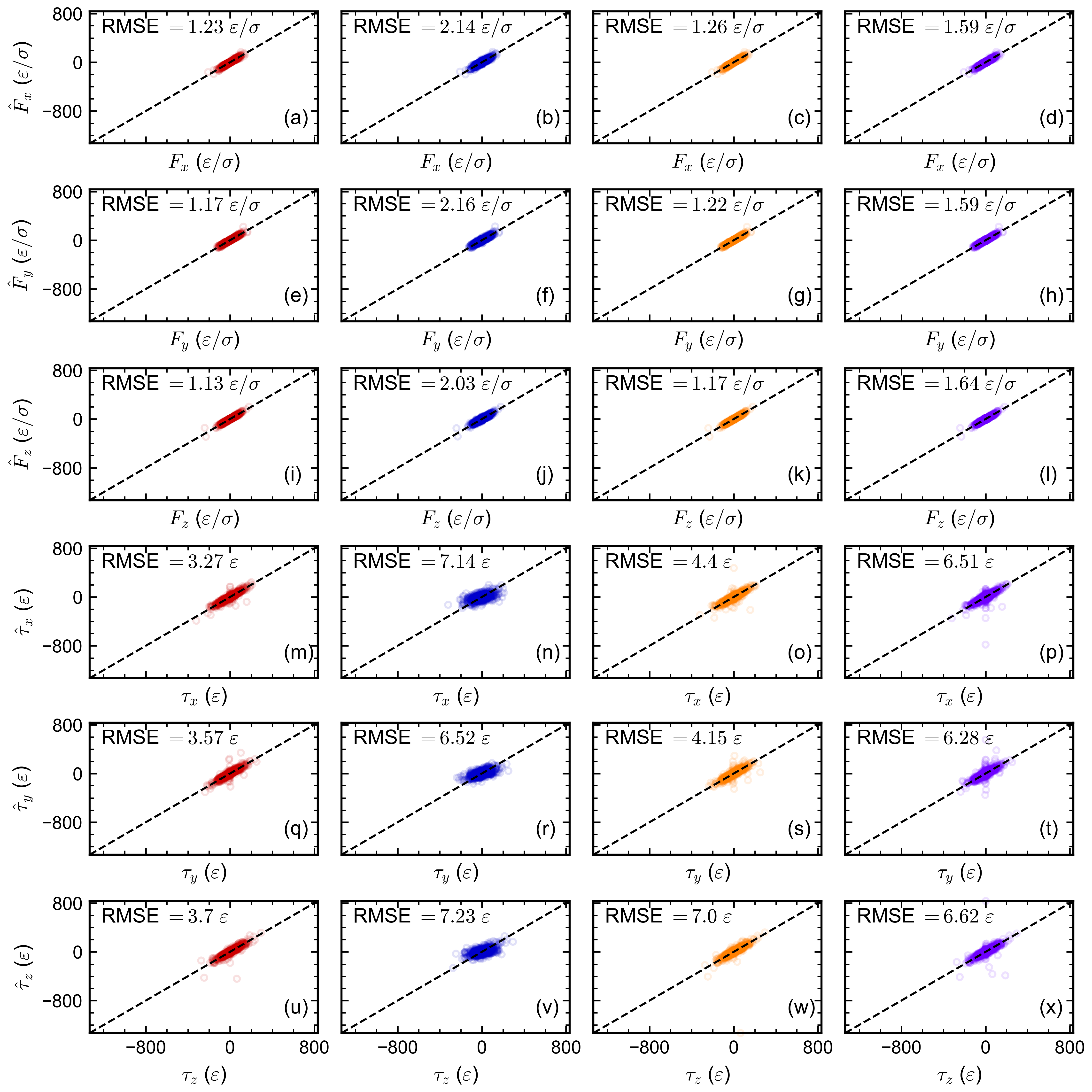}
        \caption{Same as Fig.~\ref{fig:rod3d_parity} but for the cube.}
    \label{fig:cube_parity}
\end{figure}

\begin{figure}[!h]
        \includegraphics[width=\linewidth]{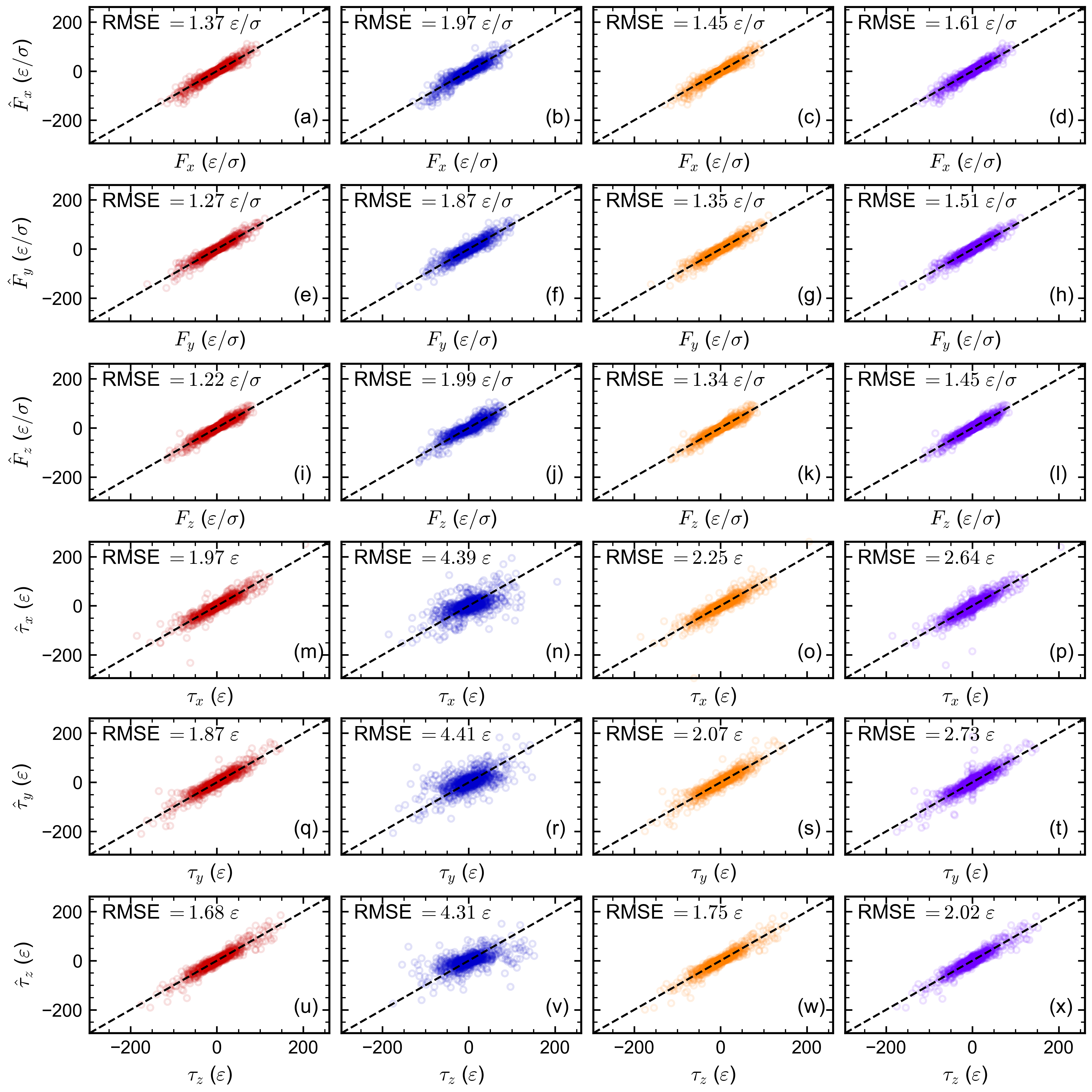}
        \caption{Same as Fig.~\ref{fig:rod3d_parity} but for the tetrahedron.}
    \label{fig:tetrahedron_parity}
\end{figure}

\begin{table}[!h]
    \caption{First (Q1), second (Q2), and third (Q3) quartile of the magnitude of curl of the forces approximated using interpolation of forces and torques (FT) and interpolation of partial derivatives (D). The units are $\varepsilon/\sigma^2$.}
    \begin{tabularx}{5in}{CCCCC}
    nanoparticle & interpolant & Q1 & Q2 & Q3 \\
    \hline
    \multirow{2}{*}{rod (2D)}  & FT & 0.02 & 0.055 & 0.169\\
    & D & 0.012 & 0.036 & 0.138\\
    \multirow{2}{*}{square}  & FT & 0.026 & 0.067 & 0.219\\
    & D & 0.02 & 0.055 & 0.204\\
    \multirow{2}{*}{triangle} & FT & 0.02 & 0.061 & 0.202\\
    & D & 0.013 & 0.042 & 0.151\\
    \multirow{2}{*}{rod (3D)}  & FT & 0.028 & 0.066 & 0.247\\
    & D & 0.014 & 0.04 & 0.219\\
    \multirow{2}{*}{cube}  & FT & 0.081 & 0.184 & 0.477\\
    & D & 0.043 & 0.093 & 0.245\\
    \multirow{2}{*}{tetrahedron} & FT & 0.175 & 0.368 & 0.778\\
    & D & 0.064 & 0.142 & 0.338\\
    \end{tabularx}
    \label{tab:curl_quartiles}
\end{table}